\documentclass[aps,prl,twocolumn,floatfix]{revtex4}

\usepackage{graphicx}
\usepackage{dcolumn}
\usepackage{bm}

\newcommand{\thetaD}{\theta_{\text{D}}}
\newcommand{\thetaE}{\theta_{\text{E}}}

\begin{document}


\title{Atom-in-jellium predictions of the shear modulus at high pressure}

\date{October 27, 2020; revisions to May 26, 2021 
   -- LLNL-JRNL-816175}

\author{Damian C. Swift}
\email{dswift@llnl.gov}
\author{Thomas Lockard}
\author{Sebastien Hamel}
\author{Christine J. Wu}
\author{Lorin X. Benedict}
\author{Philip A. Sterne}
\affiliation{%
   Lawrence Livermore National Laboratory,
   7000 East Avenue, Livermore, California 94551, USA
}

\begin{abstract}
Atom-in-jellium calculations of the Einstein frequency in condensed matter
and of the equation of state
were used to predict the variation of shear modulus from zero pressure
to $\sim 10^7$\,g/cm$^3$, for several elements relevant to 
white dwarf (WD) stars and other self-gravitating systems.
This is by far the widest range reported electronic structure calculation
of shear modulus,
spanning from ambient through the one-component plasma to
extreme relativistic conditions.
The predictions were based on a relationship between Debye temperature
and shear modulus which we assess to be accurate at the $o(10\%)$ level,
and is the first known use of atom-in-jellium theory to calculate 
a shear modulus.
We assessed the overall accuracy of the method by comparing with 
experimental measurements and more detailed electronic structure calculations
at lower pressures.
\end{abstract}


\maketitle

\section{Introduction}
The shear modulus is a fundamental measure of the resistance of matter to shear
deformation,
dictating the speed of propagation of shear waves,
contributing to the speed of longitudinal waves,
and governing the magnitude of deviatoric stresses induced by shear strains,
which are the driving force for plastic flow.
Although straightforward to measure at ambient pressure,
the shear modulus is challenging to measure at elevated pressures because
of the difficulty of distinguishing its contribution from that of the 
bulk modulus, i.e. volumetric compression of the sample.
However,
the shear modulus is a key aspect in understanding the response of solids
to deformation at high pressure, which is typically dynamic.
It represents the first-order correction to the scalar equation of state (EOS)
to account for non-hydrostatic stresses.
Technologically, the high-pressure shear modulus is important in impacts and the
response of solids to explosions, 
as occur in weapon physics and target response.
Scientifically, it occurs in planetary seismology and oscillatory modes
of white dwarf and neutron stars -- at very different pressure regimes.

The shear modulus is usually predicted theoretically from electronic structure
calculations of single-crystal elastic moduli, which are then averaged to
estimate the shear modulus of polycrystalline matter \cite{Rudd2018}.
This approach is rigorous,
but it is subject to some difficulties and inaccuracies in practice.
If the appropriate crystal structure is 
not represented accurately enough in the electronic structure model,
or contains internal degrees of freedom for which the equilibrium parameter
values are not found precisely enough,
the model of the crystal may be unstable with respect to some distortions
from the supposed equilibrium, giving unphysical negative elastic constants.
Because calculations of the elastic moduli usually break symmetries of the
equilibrium structure, and several distortions of the structure are
needed to determine the elastic moduli, the computational effort involved
is higher than for the EOS. For these reasons, predictions of elastic moduli
are generally less extensive than are EOS.

Although single-crystal elastic properties are important for some
applications, most require a polycrystalline average shear modulus.
The properties of a polycrystalline ensemble depend on the texture of the
material, which introduces another degree of freedom.
The limiting cases of Voigt and Reuss averaging -- assuming that either the
stress or strain is equal over grains of different orientation -- may be
significantly different \cite{Swift_dia_2020}.

We present a different and computationally efficient method of predicting
the shear modulus over a wide range of states, avoiding most of these
complications.
This method can be made to work with any approach to constructing
the EOS from which the ion-thermal contribution can be identified.
Here we use the atom-in-jellium method \cite{Liberman1979,Liberman1990}, 
which we have recently been investigating as a particularly efficient
approach to predicting the EOS of elements over a wide range of states
\cite{Swift_ajeos_2019,Swift_iontherm_2020}.
In fact, we have found it possible to calculate the EOS over eleven decades
in mass density and ten in temperature, the first application of
a reasonably accurate electronic structure technique to span from
ambient conditions to the core of a white dwarf star \cite{Swift_wdeos_2020}.

\section{Relation between shear modulus and ion-thermal EOS}
Although it is considered most natural to express the ion-thermal EOS of
crystalline matter in terms of phonons, there is a close connection
with the elastic moduli, as they give the frequencies of the
acoustic modes.
In the phonon approach, the thermal energy of each phonon mode has 
the Bose-Einstein form.
The ion-thermal EOS can be found by integrating over all the phonon modes
\cite{Swift_Sieos_2001}.
However, many of the phonon modes are similar, and in integrating over the
population the details of any given mode become unimportant.
It is common in constructing even recent, rigorous, multiphase EOS
to express the ion-thermal contribution as a few effective Debye modes,
or even a single mode.
Average Debye modes can be estimated from the density of phonon states,
or from the elastic moduli.
There is a one-to-one correspondence between elastic moduli and the
speeds of longitudinal and shear waves.
Depending on the details of the approach adopted, including the particular
software implementation, there can be significant advantages in
deriving the ion-thermal energy from the elastic moduli instead of
phonon modes.
Although elastic moduli are susceptible to numerical instabilities
as mentioned above, phonon modes are usually even more susceptible,
resulting in a population of imaginary modes. 
If the phonon modes are calculated by making finite displacement of ions
from equilibrium, the symmetry of the crystal lattice is often reduced
even more than by the distortions used to calculate elastic moduli.
Phonon calculations often require the electron wavefunctions to be
constructed over a supercell of the lattice, in order to reduce the effect
of image displacements in a periodic representation. 
These constraints can make phonon calculations considerably more expensive
than elastic moduli.

If the ion-thermal EOS is represented by a single Debye mode,
it is naturally related to a single shear modulus.
Compared to the calculation of the ion-thermal energy by considering
longitudinal and shear wave speeds instead of elastic moduli,
this approach is based on average wave speeds instead of the average energy,
i.e. at least in principle calculating an average over all
directions and polarizations of the elastic waves.
There is a long history of relating the Debye temperature $\thetaD$
to the shear modulus \cite{Madelung1910,Einstein1911,Anderson1963},
and this approach for predicting the ion-thermal energy is still in use
\cite{Liu2018}.
Following Anderson \cite{Anderson1963},
\begin{equation}
\thetaD(\rho)=\frac h{k_B}\left(\frac{3N_A\rho}{4\pi\bar A}\right)^{1/3}\bar u,
\end{equation}
where $h$ and $k_B$ are Planck's and Boltzmann's constants respectively,
$N_A$ is Avogadro's number, $\bar A$ is the mean atomic weight,
the average wave speed
\begin{equation}
\bar u=\left(\frac{1/u_l^3+2/u_s^3}3\right)^{1/3},
\end{equation}
and the shear and longitudinal wave speeds are
\begin{equation}
u_s=\sqrt{\frac G\rho}, u_l=\sqrt{\frac{B+4G/3}\rho}.
\end{equation}
Relating the Debye temperature and the shear modulus relies on a hierarchy
of approximations,
in this case that the material is either elastically isotropic, or 
comprises a uniform distribution of grain orientations so as to give
an isotropic average response and the shear modulus is the Hill average
\cite{Anderson1963}.

Conversely, the shear modulus $G$ may be estimated from the
Debye temperature $\thetaD$ and bulk modulus $B$.
This calculation involves inverting the function $\thetaD(B,G)$,
which we performed numerically by bracketing and bisection, defining
the bracket with factors $\beta_1$ and $\beta_2$ of $B$,
where $\beta_1\ll 1$ and $\beta_2\gg 1$.
Another approach has been to ignore either $B$ or $G$ and so make the
expression for $\thetaD$ invertible \cite{Ledbetter1991}:
$\thetaD\propto\rho^{1/3}\sqrt{G/\rho}$.
A similar cubic relation has been used to relate the shear modulus
to the elastic moduli rather than the Debye temperature, in cubic crystals
\cite{Kroener1958}.
As well as predicting the shear modulus for materials of interest
using the complete equation,
we assess the accuracy of the approximate solution.

\section{Atom-in-jellium equation of state}
The atom-in-jellium (AJ) model \cite{Liberman1979} 
of electronic structure in matter 
provides a computationally efficient way to
predict wide-ranging EOS models that provides $B$ and $\thetaD$
and is thus convenient for making wide-range predictions of the shear
modulus. 
$\thetaD$ can be estimated from a perturbative calculation of the restoring force
when the nucleus is displaced from equilibrium, which gives the Einstein
temperature $\thetaE$ \cite{Liberman1990}.
$\thetaD$ may be inferred from $\thetaE$ by equating the thermal energy
or mean squared displacement, which give significantly different values.
We found that the displacement calculation gave results which, compared with
a range of experimental data, theoretical predictions, and previous models
for a variety of elements at pressures up to $\sim 0.1$\,TPa,
were systematically lower by a factor $\sim 3$,
whereas the energy calculation was more consistent,
so we used the latter for all the results reported here.

The shear modulus is expected to be a function of both 
mass density $\rho$ and temperature $T$.
Strength models often express the shear and flow stress, $G$ and $Y$, 
in terms of pressure $p$ instead of $\rho$.
However, most materials expand with temperature along an isobar,
so $\partial G/\partial T|_\rho$ is typically smaller than
$\partial G/\partial T|_p$, and so we prefer to consider $G(\rho,T)$. 
Unusually, the AJ calculation gives $\thetaD(\rho,T)$ \cite{Swift_ajeos_2019},
so it can be used to predict the temperature-dependence of $G$ as well as the
density-dependence.
Trial calculations indicated that the variation with $T$ was not much greater
than the numerical noise in the AJ solution for $\thetaD$, and we do not
consider it further here.

In order to study the limiting trends as the atoms are compressed closely
enough for all the electrons to be unbound, the one-component plasma (OCP)
limit, we based the shear modulus calculations on AJ EOS models
of H, He, C, O, Ne, and Mg,
constructed previously for WD studies \cite{Swift_wdeos_2020}.
The WD EOS calculations were performed to a mass density $o(10^7)$\,g/cm$^3$,
which is four orders of magnitude higher than for usual general-purpose EOS
\cite{sesame,leos},
and a temperature $o(10^6)$\,eV, an order of magnitude higher than usual.
The shear moduli deduced are thus applicable at least in principle
to WDs and the crust of neutron stars.
We also calculated the shear modulus for Fe,
as an intermediate-$Z$ element of astrophysical importance
whose strength has been relatively well studied.
The Fe shear modulus was calculated from a standard-range AJ tabulation.
In all cases,
$B$ and $\thetaD$ were taken at the lowest temperature calculated in the
AJ EOS, which was 1\,K.

\begin{figure}
\begin{center}\includegraphics[scale=0.72]{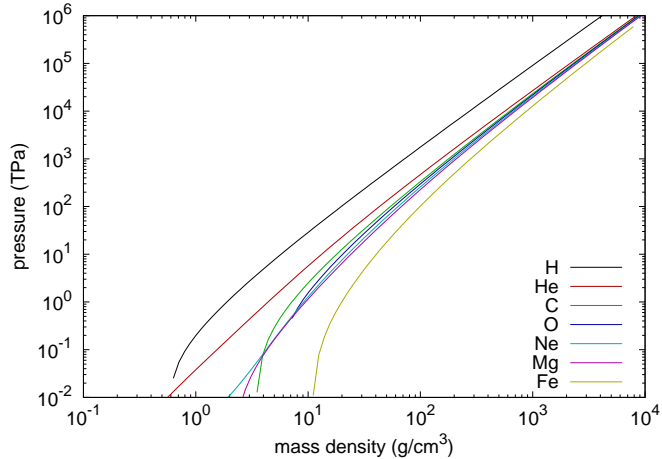}\end{center}
\caption{1\,K isotherm from AJ calculation.}
\label{fig:pcmp}
\end{figure}

\begin{figure}
\begin{center}\includegraphics[scale=0.72]{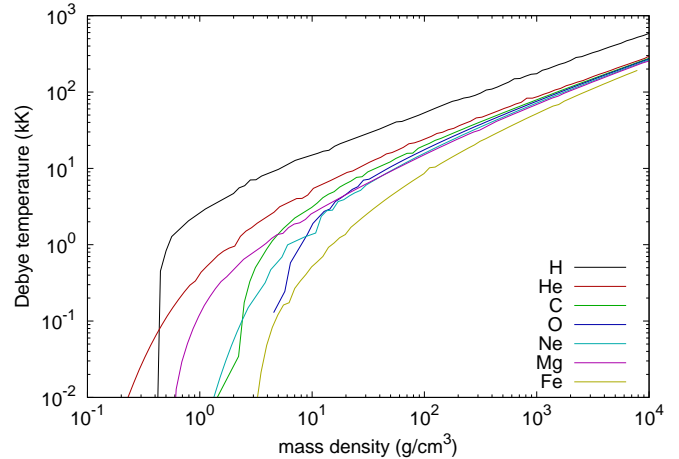}\end{center}
\caption{Debye temperature from AJ calculation
   (from Einstein temperature by equating energy).}
\label{fig:tdebcmp}
\end{figure}

The pressure and bulk modulus were found to vary as $\rho^{5/3}$ at high compression, 
as expected for the OCP, 
and then tended toward $\rho^{3/2}$ in the extreme relativistic regime.
To emphasize: these behaviors are a product of the AJ calculation, 
not imposed as an assumption.
(Figs~\ref{fig:pcmp} and \ref{fig:tdebcmp}.)

\begin{figure}
\begin{center}\includegraphics[scale=0.72]{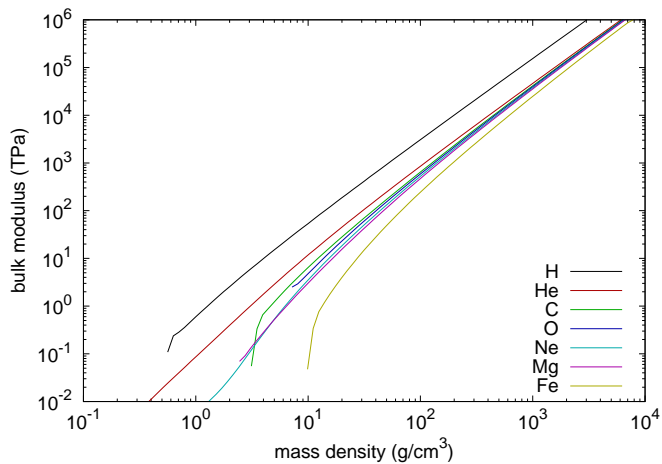}\end{center}
\caption{Bulk modulus from AJ calculation.}
\label{fig:bcmp}
\end{figure}

\begin{figure}
\begin{center}\includegraphics[scale=0.72]{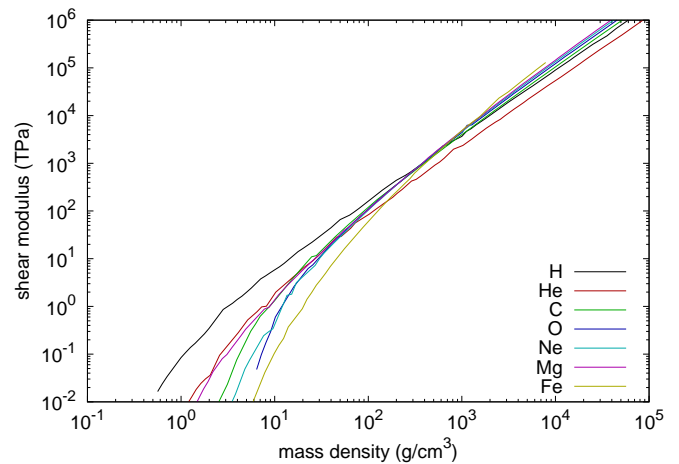}\end{center}
\caption{Summary of AJ shear modulus predictions at 1\,K.}
\label{fig:gcmp}
\end{figure}

The numerical solution of $G(B,\thetaD)$ performed reliably over the
full range of densities considered, for all elements
(Fig.~\ref{fig:gcmp}). 
The simplified calculation neglecting $B$ \cite{Ledbetter1991} gave
results typically 30\%\ higher.

The resulting $G(\rho)$ predictions were generated as tables.
For convenience, we obtained functional fits to the tabular data:
see Appendix.

It is challenging to make EOS measurements at the terapascal range and above
of materials that are solid at ambient conditions,
and even more so to measure the shear modulus.
For the elements considered here, our comparisons are primarily against
ambient measurements where available, and otherwise against other models.

\section{Previous shear modulus models}
Other models have been developed for the shear modulus at high pressure.
We summarize two of them here for which results have been published for
Mg, C, and Fe.

\subsection{Steinberg-Guinan}
The Steinberg-Guinan (SG) model \cite{Guinan1974}
is widely used in hydrodynamic simulations below 100\,GPa, and has the form
\begin{equation}
G=G_0\left[1+A p/\eta^{1/3}-B(T-T_0)\right]\quad:\quad \eta\equiv\rho/\rho_0
\end{equation}
where $p$ is the pressure.
Although developed for relatively low pressures,
it is constructed to asymptote toward the one-component plasma (OCP) limit,
where the shear modulus varies as $\rho^{4/3}$, if used in conjunction
with an EOS model in which the pressure varies as $\rho^{5/3}$.
However, the free parameters $G_0$, $A$, and $B$ are typically chosen to
match low-pressure data, and the absolute value is not constrained to
be correct in the OCP limit.

Improvements have been proposed to extend the SG model to higher pressures,
either by adopting different parameter values in high-pressure phases
\cite{Rudd2018}, or by modifying the dependence on compression
to transition to a different function at high compressions \cite{Orlikowski2007}:
\begin{eqnarray}
G&=&G_0\left\{f(\rho)G_l+\left[1-f(\rho)\right]G_h-B(T-T_0)\right\} \\
f(\rho)&=&\left\{1+\exp\left[\alpha(\eta(\rho)-\eta_c\right]\right\}^{-1} \\
G_l&=&1+A_l p/\eta(\rho)^{1/3} \\
G_h&=&A_h+M_h\eta(\rho) \\
\eta(\rho)&=&\rho/\rho_0.
\end{eqnarray}
Typically, the high pressure term $G_h$ is calibrated 
against electronic structure calculations that extend into the terapascal range,
but do not explore the OCP limit.
Ironically, it is the low pressure term $G_l$ 
that asymptotes to the expected OCP behavior,
but is masked by the softer, linear dependence on $\rho$ in $G_h$.

As the SG and improved SG (ISG) models depend on the pressure as well as
the mass density, an EOS is needed when calculating the shear modulus.
For consistency across all models, we used the AJ EOS.
The AJ method is typically less accurate at pressures below a few tenths
of a terapascal, so other models may appear to be
less accurate at low pressures than with alternative EOS.

\subsection{Straub}
In theoretical studies using early electronic structure predictions,
it was observed that the variation of shear modulus $G$
with the lattice parameter $a$ in cubic crystals such as W
behaves similarly to the bulk modulus $B(a)$.
A similar form of fitting function was adopted as was used for 
the cold curve energy $E(a)$,
\begin{equation}
G=G_0+g_a\frac{a-a_0}{a^2}e^{-g_2(a-a_0)}.
\end{equation}
Parameters were fitted to electronic structure datapoints $\{a_i,G_i\}$ 
or to $G_0$ and $d\,\ln\,G/d\,\ln\,a$
\cite{Straub1990}.
Shear moduli have been predicted in this way for a small number of elements,
and included as tabulations of $G(\rho)$ in 
the {\sc sesame} library of material properties \cite{sesame}. 


\section{Discussion}
The AJ method is known to be inaccurate at low pressure in comparison
with multi-ion electronic structure techniques, and the derived calculation
of shear modulus involves unquantified approximations.
In particular, the AJ method does not account for angular forces such as occur
in molecular bonds.
It is interesting to compare with a recent analysis invoking the shear
contribution to the longitudinal sound speed $c_l$ in atomic matter, 
and comparing with
high-fidelity electronic structure calculations of H \cite{Trachenko2020}.
$c_l$ calculated from the AJ shear and bulk moduli
agrees very well at low pressures with the theoretical analysis, to which
multi-ion electronic structure calculations asymptote as H$_2$ molecules
dissociate on compression.
However, we find that $c_l$ is dominated by the contribution from $B$
for H in this regime, $G$ being $\sim 50$ times smaller than $B$.

\begin{figure}
\begin{center}\includegraphics[scale=0.72]{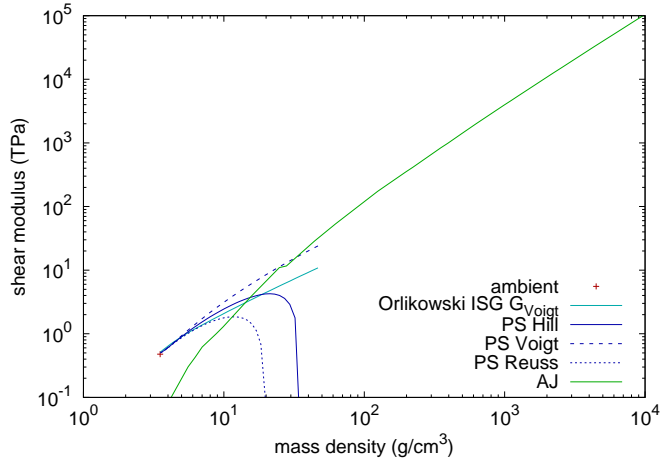}\end{center}
\caption{Comparison of AJ shear modulus prediction with observed and
   calculated behavior for C,
   including pseudopotential based multi-ion calculations (PS)
   \cite{Swift_dia_2020} and ISG model \cite{Orlikowski2007}.
   Pseudopotential predictions demonstrate possible range of shear modulus
   related to texture, and localized reductions in shear modulus
   associated with mechanical instability at a phase transition.
   }
\label{fig:c}
\end{figure}

For C, the AJ shear modulus falls well below the observed value for
diamond at STP, which is not surprising for a structure stabilized by
directional bonding.
The prediction passes through our recent pseudopotential predictions
\cite{Swift_dia_2020}
above 10\,g/cm$^3$, crossing the Hill average just below 20\,g/cm$^3$ where
the diamond structure was predicted to start to become unstable,
and at higher is more consistent with an extrapolation of the
Voigt polycrystal average in diamond.
(Fig.~\ref{fig:c}).

\begin{figure}
\begin{center}\includegraphics[scale=0.72]{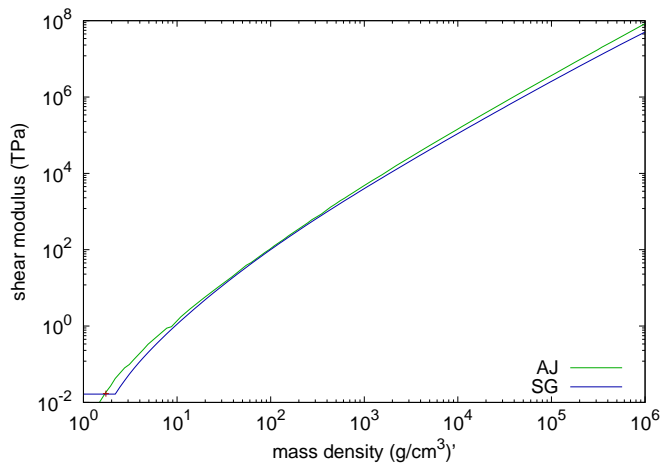}\end{center}
\caption{Comparison of AJ shear modulus prediction for Mg with STP measurement
   and SG model.}
\label{fig:mg}
\end{figure}

For Mg, which adopts the hexagonal close-packed structure at low pressures, 
the AJ shear modulus reproduces the observed STP value to within a few percent.
It follows the SG model quite closely over a wide pressure range.
The SG model appears to have a discrepancy at low pressures;
as discussed above,
this is an artifact caused by using the AJ EOS to calculate the
pressure, and is an example of better performance of electronic structure
calculations in predicting derivatives of pressure than for the pressure 
itself \cite{Akbarzadeh1993}.
The AJ prediction becomes quite close to the SG model 
above 10\,g/cm$^3$ ($\sim$1\,TPa),
and as the compression increases further, the AJ shear modulus gradually
becomes a few tens of percent stiffer than the SG. 
The SG model in this regime is constrained only by its asymptotic form,
and this result is an example of the SG model performing remarkably well.
Mg exhibits solid-solid phase transitions \cite{Pickard2010,Li2010},
but the observed and predicted structures are close-packed structures
or perturbations of simple structures stabilized by interactions between
inner electrons at high pressures,
and the AJ calculation is likely to be reasonable for bulk average
mechanical properties.
(Fig.~\ref{fig:mg}).

\begin{figure}
\begin{center}\includegraphics[scale=0.72]{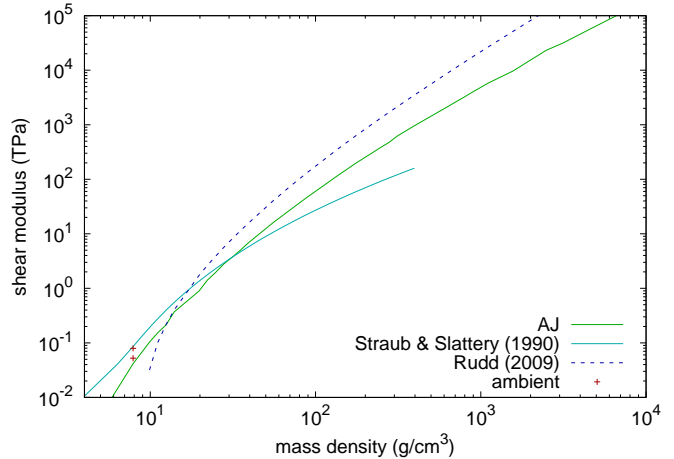}\end{center}
\caption{Comparison of AJ shear modulus prediction for Fe with STP measurements
   and models, including the Straub model which is {\sc sesame} table 32140
   \cite{ses32140} and the ISG model derived from multi-ion
   electronic structure calculations to the hundred gigapascal regime
   \cite{Rudd2009}.}
\label{fig:fe}
\end{figure}

Fe exhibits solid phases, the low-pressure bcc structure being stabilized by
magnetism, which the AJ model does not capture.
The shear modulus of Fe at STP depends on the C content: low-C steels tend to
be less rigid, and the AJ shear modulus lies relatively close to the
lower reported values at STP.
The Straub model is more consistent with C-rich Fe at low pressures,
then passes through the AJ calculation around 4\,TPa and lies well below
at higher pressures.
A SG calibration has been made for Fe at pressures of a few hundred gigapascals
using multi-ion electronic structure calculations \cite{Rudd2009}.
The AJ shear modulus intersects this model around 200\,GPa and lies well below
it at higher pressures.
We suggest that these comparisons on balance favor the AJ calculation over a 
wide range of pressures.
(Fig.~\ref{fig:fe}).

Shear modulus predictions based on multi-ion electronic structure
are more accurate in principle than the AJ predictions presented here.
However, in practice, differences in technique and the
need to calculate and combine multiple elastic moduli means that larger
variations are commonly seen \cite{Swift_dia_2020},
and it has been difficult in practice to generate wide-range predictions
of shear modulus.
Also, multi-ion calculations must be performed in a mechanically-stable phase,
involving extra effort to identify an appropriate phase at each state.
Dynamic loading experiments have commonly been performed outside the range of
meaningful estimates of the shear modulus: this is no longer necessary for
elements \cite{Swift_ajsheardata_2020}.
In high-pressure experiments, material strength is often a small effect
in response dominated by the scalar EOS, and uncertainties
$\sim 10\%$ in shear modulus appear correspondingly smaller in the overall
response of the material.
Because most materials are elastically anisotropic at some level,
the effective shear modulus depends on the microstructural texture of the
material, which may evolve during dynamic loading.
Compared with microstructural effects (Fig.~\ref{fig:c}), 
an uncertainty $\sim 10\%$ is potentially insignificant.

\section{Conclusions}
Shear moduli were predicted for condensed matter from zero pressure to
beyond the OCP regime using predictions of the bulk modulus and $\thetaD$
from AJ theory.
Although the predicted shear moduli are likely to be inaccurate
for crystal structures stabilized by angular forces, which are not
captured by AJ theory, they appear to be a reasonable choice over a 
wide range of compressions when a more rigorous model is not available.
The likely accuracy of the shear modulus predictions reflects the
uncertainties in the underlying methods: the approach adopted here could be
used with more accurate treatments of electronic structure when the 
corresponding calculations of elastic moduli are not available, for
instance for alloys and compounds.

Electronic structure calculations using the AJ method are
often inaccurate around zero pressure, 
but appear to be accurate above a few hundred gigapascals,
depending on the element.
The method is valid to extreme relativistic conditions -- beyond the OCP regime
-- of density and temperature.
AJ does not capture crystal structures and directional bonding: 
it is likely to be inaccurate in low-pressure structures or 
around phase transitions, maybe by $\sim$100\%.
As with other pressure derivatives, the shear modulus otherwise seems to be 
predicted more accurately than the absolute pressure.
Aside from numerical noise in the nuclear perturbation calculation,
the calculation of $\thetaD$ is an approximate average.
The inaccuracy may be of order 10\%, though predicted trends
are likely to be better.


The likely performance correlates with the crystal structure.
Non-close-packed structures at low pressure
are represented poorly in the AJ electron model, 
so the ion model and EOS are likely to be inaccurate.
AJ typically fails to predict bound matter
at the observed zero-pressure density.
The shear modulus is then also likely to be inaccurate,
except for fortuitous cancellations of error.
Close-packed structures are captured reasonably
in the AJ electronic model, 
particularly at elevated pressure, so the ion model and EOS are likely
to be reasonable, as is the shear modulus.
As for the bulk modulus, the shear modulus seems to be predicted 
more accurately than the absolute pressure.
The performance is probably similar for amorphous and glassy structures.
For lower-symmetry structures at high pressure, when these are
perturbations to, or stacking faults in, close-packed structures,
the quality of shear modulus predictions is likely to be similar 
to that for the close-packed structures.
Open structures stabilized by strong directional bonds are likely to be 
less accurate.
In unstable and mixed phases, the shear modulus may be small, 
which is not captured in the AJ predictions. 
The performance should not however be affected by whether a structure is
metastable or not.

We have also developed a functional form capable of representing the 
AJ shear moduli over a wide range, although it is not valid over the
full range of the AJ calculations.

\section*{Acknowledgments}
This work was performed under the auspices of
the U.S. Department of Energy under contract DE-AC52-07NA27344.


\section{Appendix: Fit to AJ predictions}
We tried using existing strength models to fit the AJ shear modulus data,
but did not manage to find parameter sets valid over the wide range of density
of the AJ calculations.
This is not to claim that reasonable fits are impossible to find, but
fitting involves iterative optimization of parameters with a non-linear
dependence on the goodness of fit, which are often susceptible to numerical
problems.
A more general structure of model might involve a set of somewhat different
functional forms valid over restricted ranges of density, with a transition
function between each, as in the ISG model for diamond \cite{Orlikowski2007}.
We were unable to determine significant parameter values for a separate
power dependence from the AJ shear modulus predictions at low density.
Instead, reasonable fits were obtained 
using a single value at the reference density
and the switching function itself to describe stiffening at low compression:
\begin{equation}
G(\rho)=G_0 f(\rho)+G_1\left(\frac\rho{\rho_0}\right)^{p_1}\left[1-f(\rho) \right]
\end{equation}
where
\begin{equation}
f(\rho)=\exp\left(-\frac{\mu(\rho)}{\mu_t}\right)\quad :\quad
\mu(\rho)\equiv\frac \rho{\rho_0}-1
\end{equation}
and $G_0$, $G_1$, $p_1$ and $\mu_t$ are parameters.
This functional form does not capture the AJ predictions in expansion,
but this region is explored little in practice as materials spall in tension,
limiting the distension of the bulk material.
In general, this functional form does not represent the AJ shear modulus
to the OCP regime with satisfactory accuracy at intermediate compressions,
so the exponent $p_1$ was included as a parameter in a finite-range fit.
(Table~\ref{tab:ajfit}.)

\begin{table*}
\caption{Fitting parameters for AJ shear modulus.}
\label{tab:ajfit}
\begin{center}
\begin{tabular}{|l|r|r|r|r|r|l|}\hline
 & $\rho_0$ & $G_0$ & $\mu_t$ & $G_1$ & $p_1$ & Notes \\
 & (g/cm$^3$) & (GPa) & & (GPa) & & \\ \hline
H & 0.5 & 14.5 & 3.0 & 82 & 1.43 & $\ge\rho_0$, which is arbitrary, and $\le$500\,g/cm$^3$. \\
H & 0.5 & 15 & 3.9 & 103 & 1.379 & $\ge\rho_0$, which is arbitrary, and $\le 5\times 10^5$\,g/cm$^3$. \\
He & 0.5 & 0.3 & 26 & 55 & 1.390 & $\ge$2\,g/cm$^3$. $\rho_0$ is arbitrary. $G_0$ is consistent with zero. \\
C & 3.52 & 280 & 13 & $1.64\times 10^3$ & 1.378 & $\ge$7\,g/cm$^3$ \\
O & 6 & 24 & 11.6 & $3.6\times 10^3$ & 1.392 & $\ge\rho_0$, which is arbitrary. \\
Ne & 4.8 & 900 & 22 & $2.7\times 10^3$ & 1.40 & $\ge$20\,g/cm$^3$. $\rho_0$ is arbitrary. \\
Ne & 4.8 & 600 & 7.3 & $1.0\times 10^3$ & 1.57 & for 12 to 5000\,g/cm$^3$. $\rho_0$ is arbitrary. \\
Mg & 1.738 & $3.0\times 10^4$ & 140 & $8.0\times 10^2$ & 1.39 & $\ge$100\,g/cm$^3$ \\
Mg & 1.738 & 19 & 1.8 & 87 & 1.73 & for 3 to 1000\,g/cm$^3$ \\
Fe & 7.874 & 52.5 & 9.3 & $1.2\times 10^3$ & 1.68 & \\
\hline\end{tabular}
\end{center}
\end{table*}

Where possible, the STP values of $\rho$ and $G$ were used as parameters, 
and low pressure AJ points were deweighted or removed if necessary. 
Otherwise, such as where AJ fails to capture solid phases 
with a significantly different shear modulus, 
the AJ data were fitted as far down in pressure as possible, 
$G_0$ was fitted if necessary, 
and $\rho_0$ was also adjusted if needed to keep $G_0>0$. 
The resulting models are not intended for use at low pressure, 
though some are probably adequate for practical purposes.

\begin{figure}
\begin{center}\includegraphics[scale=0.72]{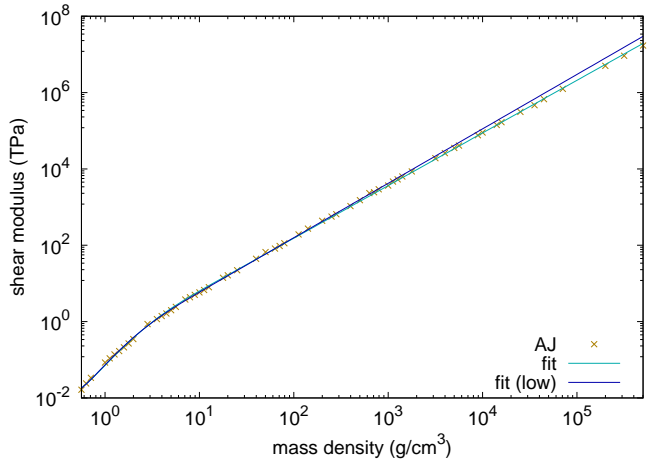}\end{center}
\caption{Comparison of AJ shear modulus predictions for H with fits.}
\label{fig:H_fit}
\end{figure}

\begin{figure}
\begin{center}\includegraphics[scale=0.72]{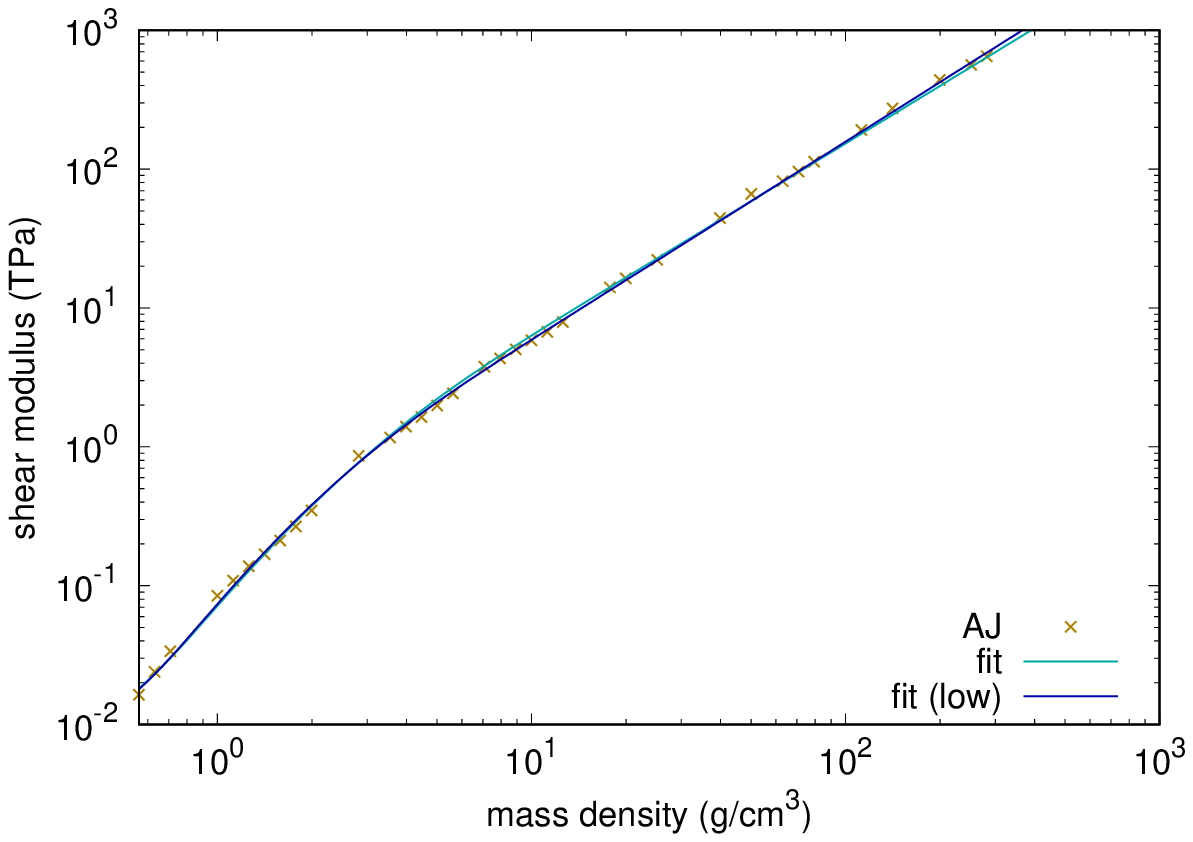}\end{center}
\caption{Comparison of AJ shear modulus predictions for H with fits
   (detail at lower compression).}
\label{fig:H_fitlo}
\end{figure}

\begin{figure}
\begin{center}\includegraphics[scale=0.72]{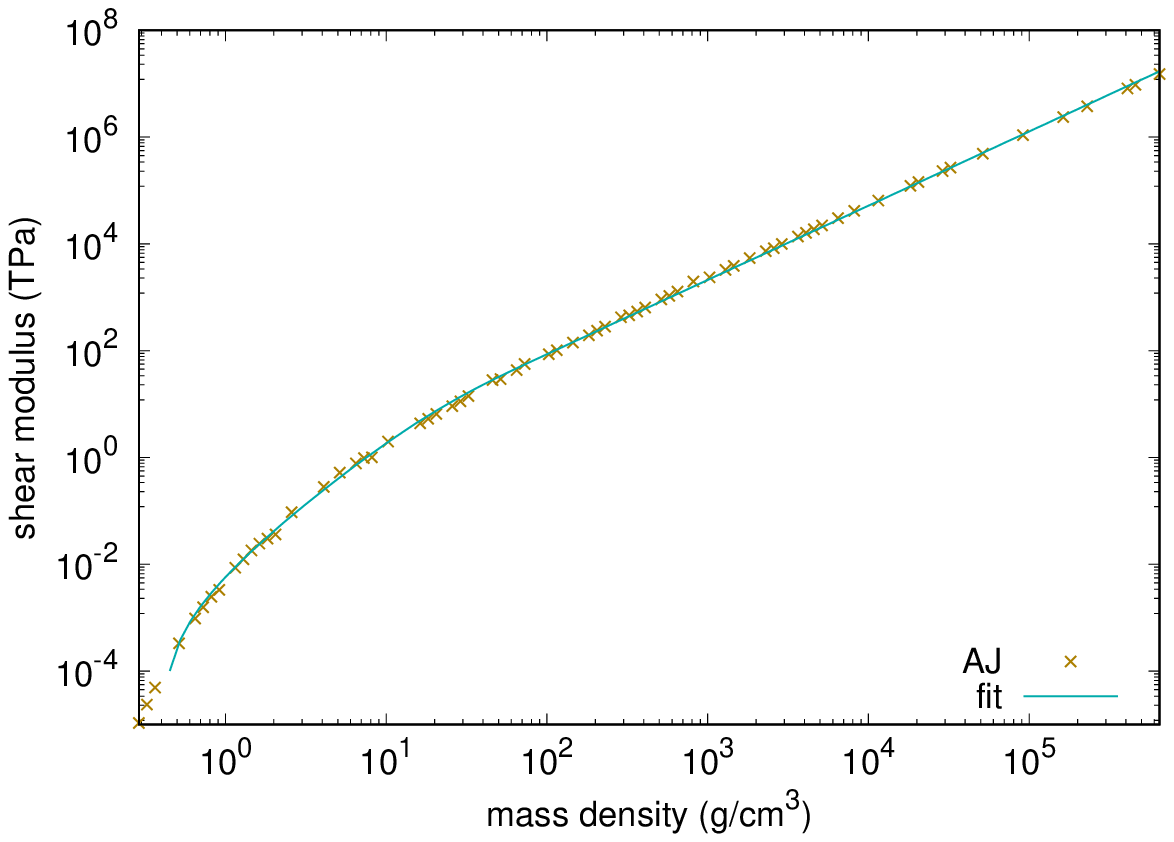}\end{center}
\caption{Comparison of AJ shear modulus predictions for He with fit.}
\label{fig:He_fit}
\end{figure}

\begin{figure}
\begin{center}\includegraphics[scale=0.72]{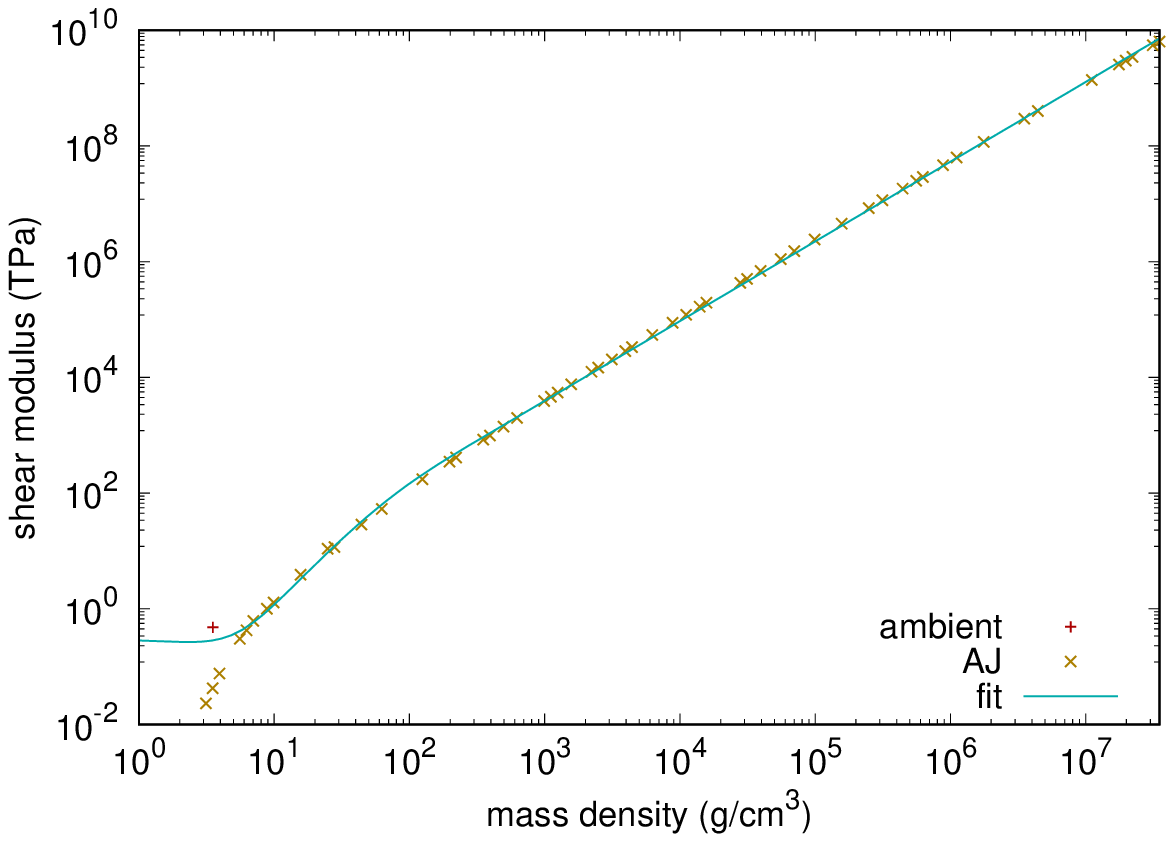}\end{center}
\caption{Comparison of AJ shear modulus predictions for C with fit.}
\label{fig:C_fit}
\end{figure}

\begin{figure}
\begin{center}\includegraphics[scale=0.72]{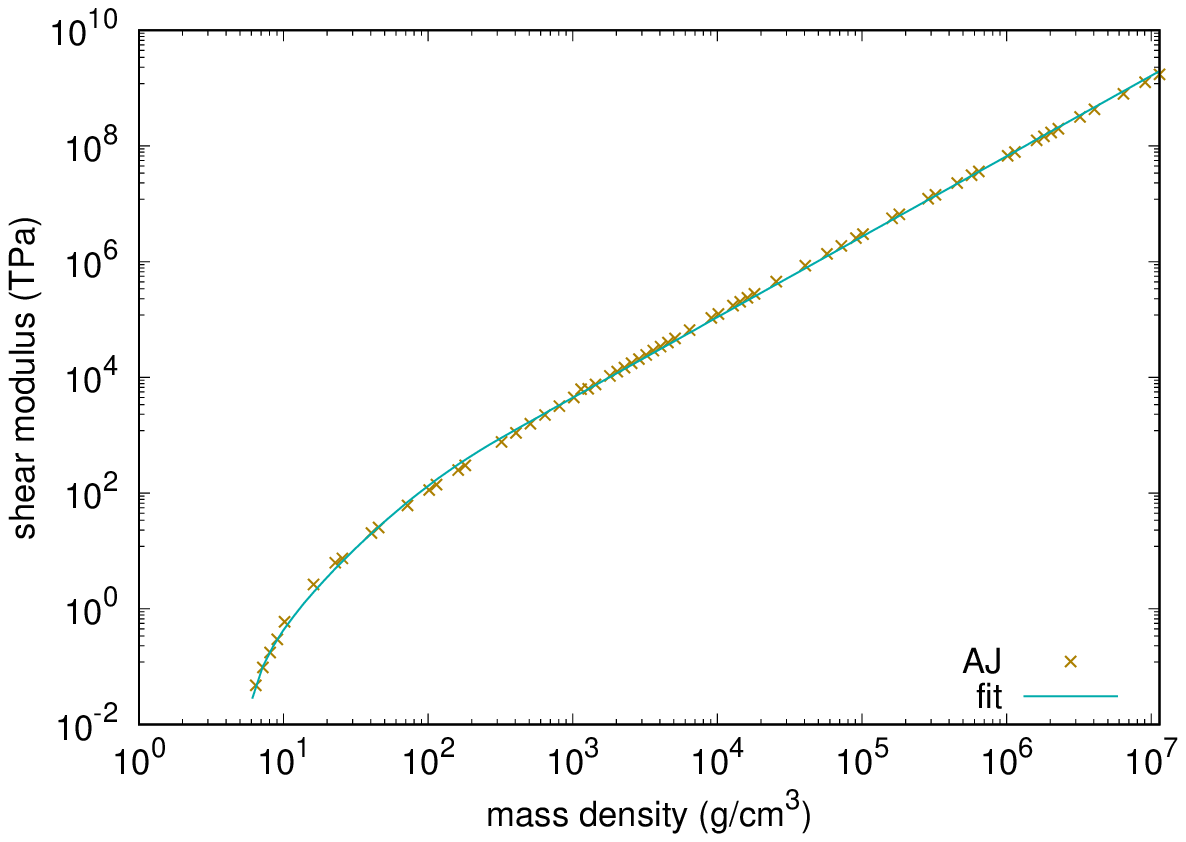}\end{center}
\caption{Comparison of AJ shear modulus predictions for O with fit.}
\label{fig:O_fit}
\end{figure}

\begin{figure}
\begin{center}\includegraphics[scale=0.72]{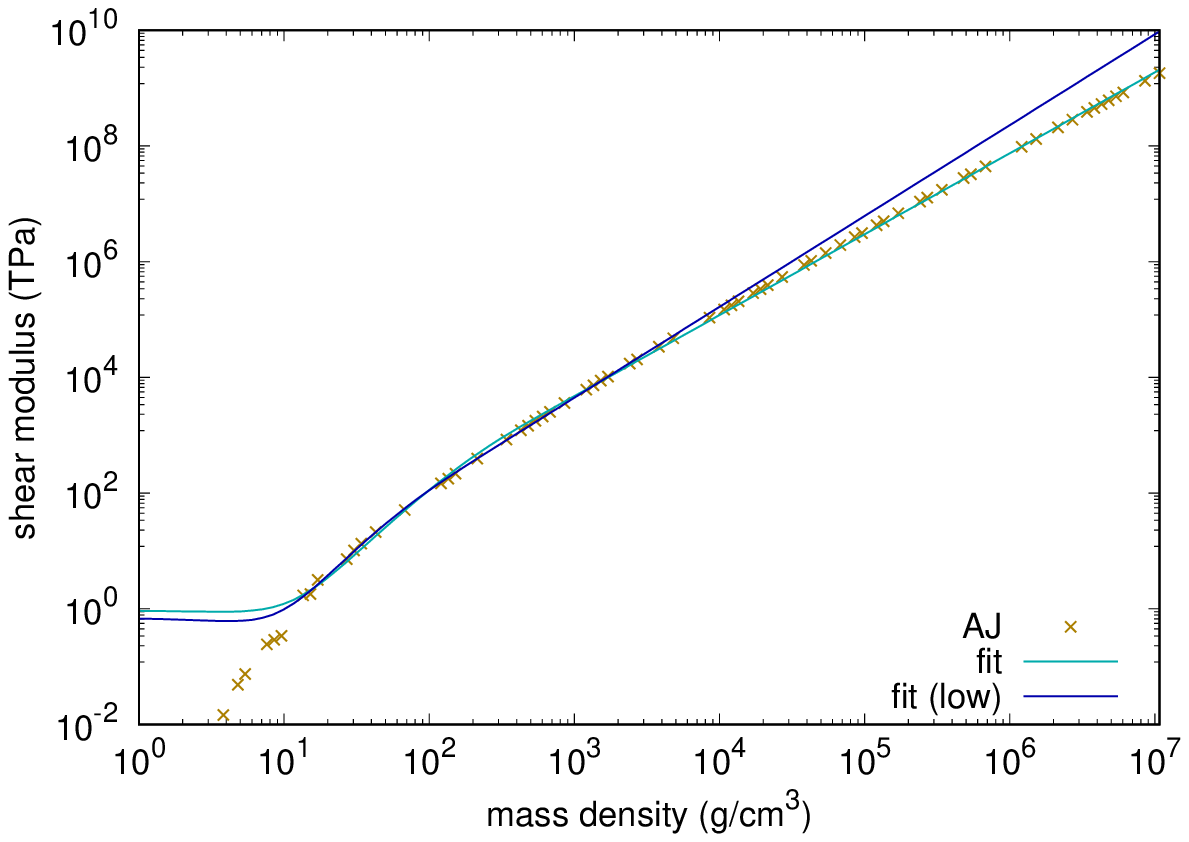}\end{center}
\caption{Comparison of AJ shear modulus predictions for Ne with fit.}
\label{fig:Ne_fit}
\end{figure}

\begin{figure}
\begin{center}\includegraphics[scale=0.72]{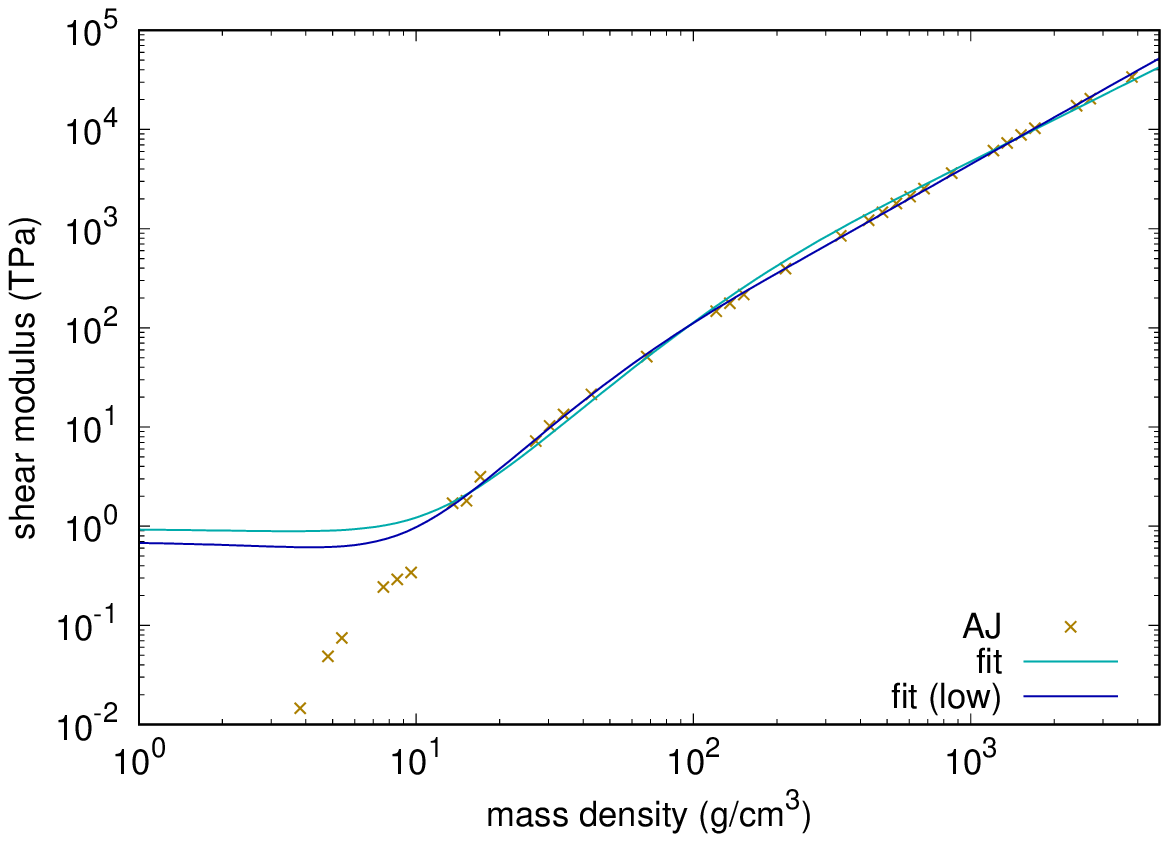}\end{center}
\caption{Comparison of AJ shear modulus predictions for Ne with fits
   (detail at lower compression).}
\label{fig:Ne_fitlo}
\end{figure}

\begin{figure}
\begin{center}\includegraphics[scale=0.72]{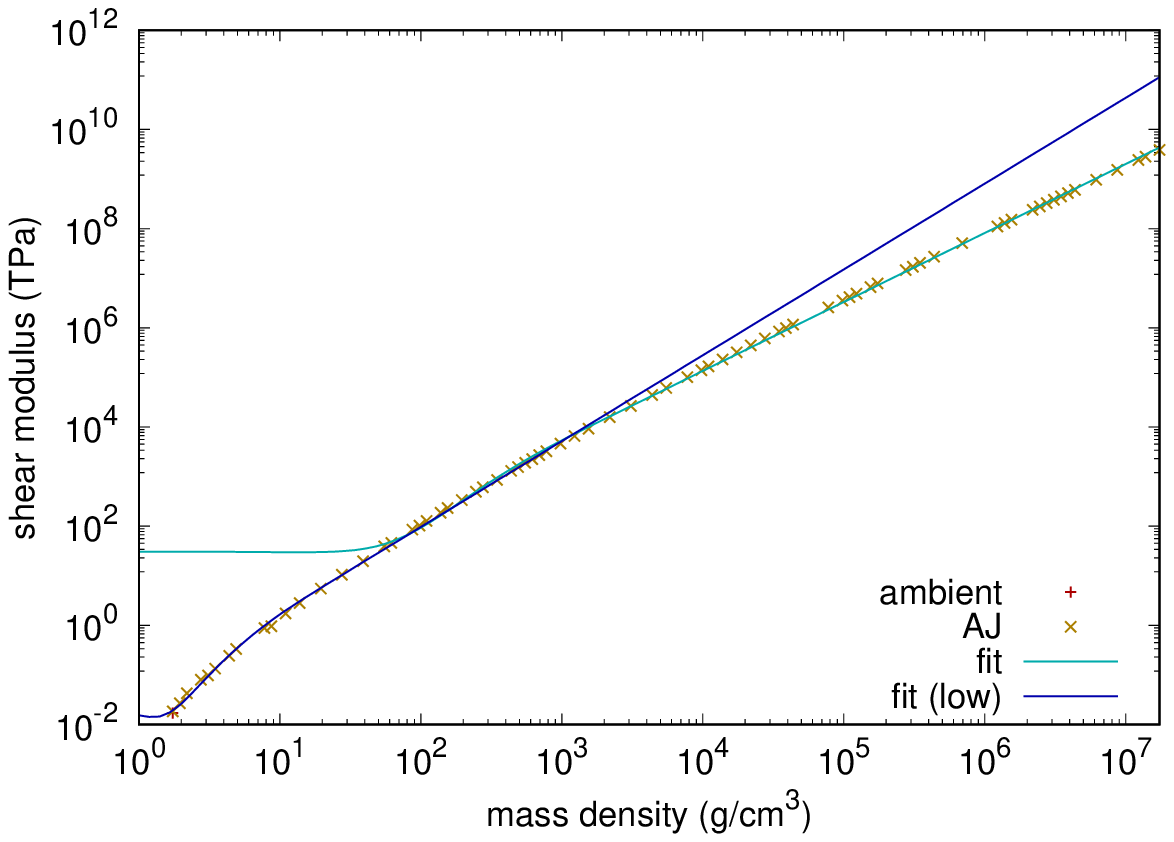}\end{center}
\caption{Comparison of AJ shear modulus predictions for Mg with fits.}
\label{fig:Mg_fit}
\end{figure}

\begin{figure}
\begin{center}\includegraphics[scale=0.72]{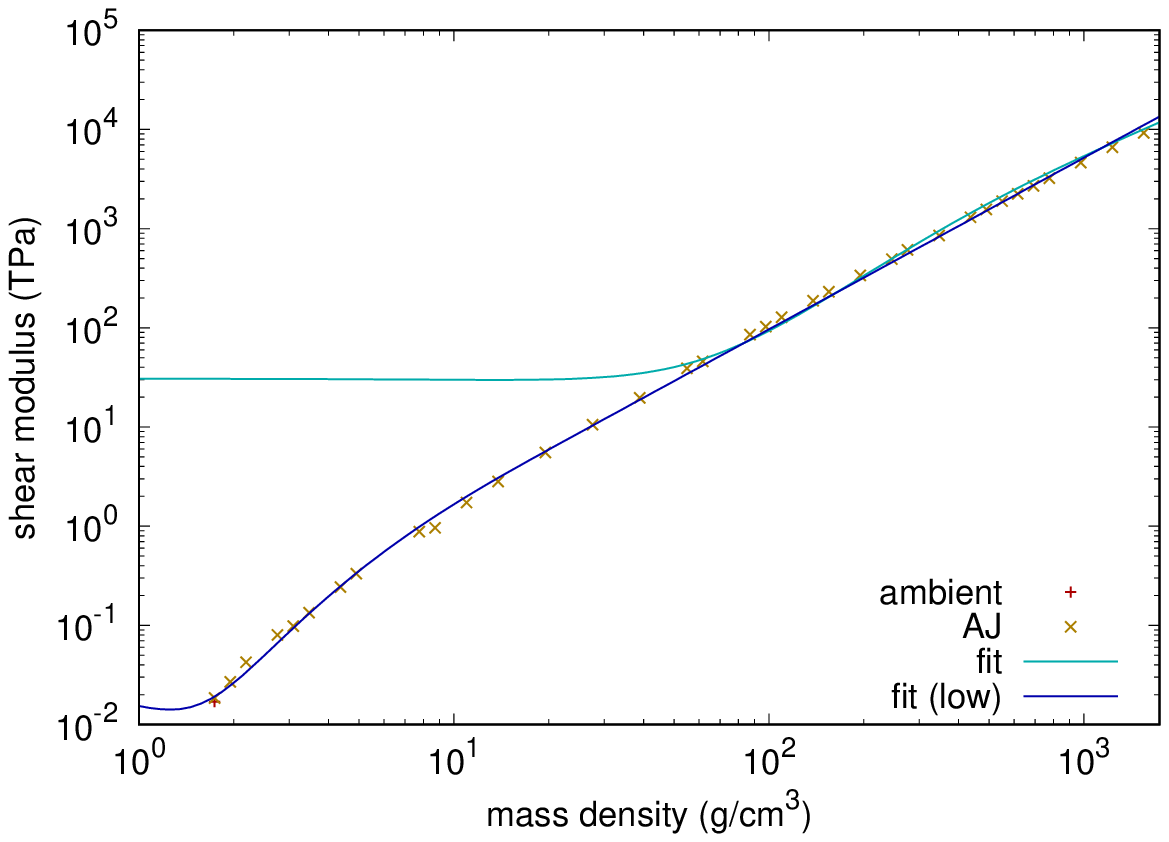}\end{center}
\caption{Comparison of AJ shear modulus predictions for Mg with fits
   (detail at lower compression).}
\label{fig:Mg_fitlo}
\end{figure}

\begin{figure}
\begin{center}\includegraphics[scale=0.72]{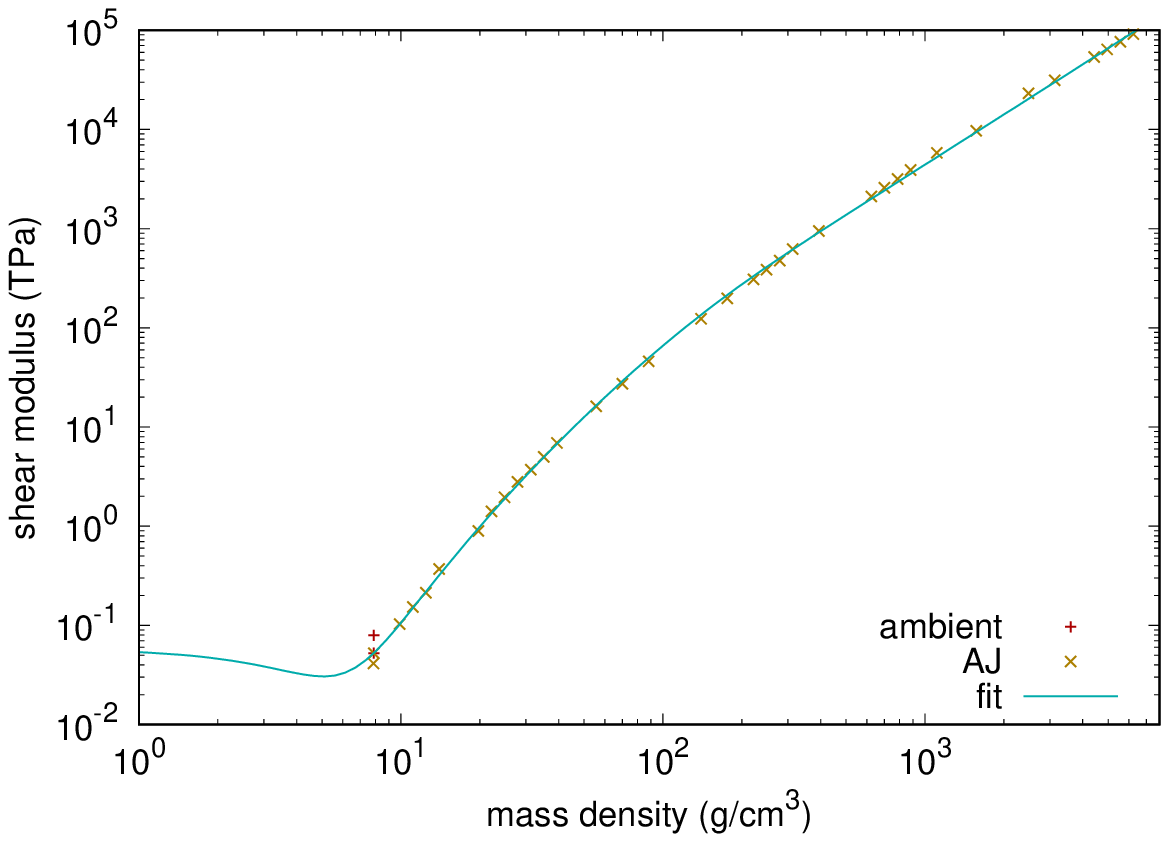}\end{center}
\caption{Comparison of AJ shear modulus predictions for Fe with fit.}
\label{fig:Fe_fit}
\end{figure}

The fitted equation usually matches the AJ data to within a few percent.
Between numerical noise in the AJ calculation
and probably-physical structure not captured by the equation,
the deviation could be up to 20\%\ in some places in most models, 
and 30\%\ in a few
(Figs~\ref{fig:H_fit} to \ref{fig:Fe_fit}).

\end{document}